\documentclass[conference,letterpaper,10pt]{IEEEtran}
\IEEEoverridecommandlockouts

\usepackage{graphicx}
\usepackage{amsmath,amssymb,mathtools}
\usepackage{booktabs}
\usepackage{multirow}
\usepackage{makecell}
\usepackage{xcolor}
\usepackage{cite}
\usepackage{bbm} 
\usepackage[caption=false,font=footnotesize]{subfig}
\usepackage[hidelinks]{hyperref}

\usepackage[acronyms,nonumberlist,nopostdot,nomain,nogroupskip,acronymlists={hidden}]{glossaries}

\newacronym{3gpp}{3GPP}{3rd Generation Partnership Project}
\newacronym{4g}{4G}{4th generation}
\newacronym{5g}{5G}{5th generation}
\newacronym{6g}{6G}{6th generation}
\newacronym{5gc}{5GC}{5G Core}
\newacronym{adc}{ADC}{Analog to Digital Converter}
\newacronym{JEPA}{JEPA}{Joint embedding predictive architecture}
\newacronym{ai}{AI}{Artificial Intelligence}
\newacronym{LE}{LE}{Log-Euclidean}
\newacronym{TP}{TP}{Test Point}
\newacronym{BLE}{BLE}{Bluetooth low energy}

\newacronym{UWB}{UWB}{Ultra-Wideband}
\newacronym{SSL}{SSL}{self-supervised learning}
\newacronym{EMS}{EMS}{electromagnetic skin}
\newacronym{mmW}{mmW}{millimeter wave}
\newacronym{FTM}{FTM}{Fine Timing Measurement}
\newacronym{AoA}{AoA}{Angle of Arrival}
\newacronym{AoD}{AoD}{Angle of Departure}
\newacronym{AoR}{AoR}{Angle of Reflection}
\newacronym{AoI}{AoI}{Angle of Incidence}
\newacronym{RIS}{RIS}{Reconfigurable Intelligent Surface}
\newacronym{RISs}{RISs}{Reconfigurable Intelligent Surfaces}
\newacronym{SNR}{SNR}{signal-to-noise ratio}
\newacronym{AF}{AF}{amplify-and-forward}
\newacronym{DF}{DF}{decode-and-forward}
\newacronym{STAR-RISs}{STAR-RISs}{Simultaneous Transmit and Reflecting RISs}
\newacronym{3GPP}{3GPP}{3rd Generation Partnership Project}
\newacronym{RAN}{RAN}{Radio Access Network}
\newacronym{KS}{KS}{Kruskal Stress}
\newacronym{CT}{CT}{Continuity}
\newacronym{TW}{TW}{Trustworthiness }
\newacronym{NCR}{NCR}{Network-Controlled Repeater}
\newacronym{NCRs}{NCRs}{Network-Controlled Repeaters}
\newacronym{IAB}{IAB}{Integated-Access-and-Backhauling}
\newacronym{UAV}{UAV}{Unmanned Aerial Vehicle}
\newacronym{SRE}{SRE}{Smart Radio Environment}
\newacronym{ecdf}{ECDF}{Empirical Cumulative Distribution Function}
\newacronym{CDF}{CDF}{Cumulative Distribution Function}
\newacronym{HSRE}{HSRE}{Heterogeneous SRE}
\newacronym{aimd}{AIMD}{Additive Increase Multiplicative Decrease}
\newacronym{am}{AM}{Acknowledged Mode}
\newacronym{GSTC}{GSTC}{Generalized Sheet Transition Condition}
\newacronym{amc}{AMC}{Adaptive Modulation and Coding}
\newacronym{FCAE}{FCAE}{Fully Connected Autoencoder}
\newacronym{amf}{AMF}{Access and Mobility Management Function}
\newacronym{aops}{AOPS}{Adaptive Order Prediction Scheduling}
\newacronym{api}{API}{Application Programming Interface}
\newacronym{apn}{APN}{Access Point Name}
\newacronym{LoS}{LoS}{Line-of-Sight}
\newacronym{NLoS}{NLoS}{None-Line-of-Sight}
\newacronym{NLDR}{NLDR}{Nonlinear Dimensionality Reduction}
\newacronym{NNs}{NNs}{Neural networks}

\newacronym{ap}{AP}{Access Point}
\newacronym{ae}{AE}{Autoencoder}
\newacronym{aqm}{AQM}{Active Queue Management}
\newacronym{ar}{AR}{Augmented Reality}
\newacronym{6G}{6G}{sixth-generation}
\newacronym{MLE}{MLE}{Mean Localization Error}
\newacronym{ausf}{AUSF}{Authentication Server Function}
\newacronym{avc}{AVC}{Advanced Video Coding}
\newacronym{awgn}{AGWN}{Additive White Gaussian Noise}
\newacronym{balia}{BALIA}{Balanced Link Adaptation Algorithm}
\newacronym{bbu}{BBU}{Base Band Unit}
\newacronym{bdp}{BDP}{Bandwidth-Delay Product}
\newacronym{ber}{BER}{Bit Error Rate}
\newacronym{bf}{BF}{Beamforming}
\newacronym{bler}{BLER}{Block Error Rate}
\newacronym{OSM}{OSM}{Open Street Map}
\newacronym{MSE}{MSE}{Mean Square Error}
\newacronym{brr}{BRR}{Bayesian Ridge Regressor}
\newacronym{bs}{BS}{Base Station}
\newacronym{bsr}{BSR}{Buffer Status Report}
\newacronym{bss}{BSS}{Business Support System}
\newacronym{ca}{CA}{Carrier Aggregation}
\newacronym{caas}{CaaS}{Connectivity-as-a-Service}
\newacronym{GPS}{GPS}{Global Positioning System}
\newacronym{cb}{CB}{Code Block}
\newacronym{cc}{CC}{channel charting}
\newacronym{ccid}{CCID}{Congestion Control ID}
\newacronym{cco}{CC}{Carrier Component}
\newacronym{cdd}{CDD}{Cyclic Delay Diversity}

\newacronym{cdn}{CDN}{Content Distribution Network}
\newacronym{cn}{CN}{Core Network}
\newacronym{codel}{CoDel}{Controlled Delay Management}
\newacronym{comac}{COMAC}{Converged Multi-Access and Core}
\newacronym{cord}{CORD}{Central Office Re-architected as a Datacenter}
\newacronym{cornet}{CORNET}{COgnitive Radio NETwork}
\newacronym{cosmos}{COSMOS}{Cloud Enhanced Open Software Defined Mobile Wireless Testbed for City-Scale Deployment}
\newacronym{cots}{COTS}{Commercial Off-the-Shelf}
\newacronym{cp}{CP}{Control Plane}
\newacronym{cyp}{CP}{Cyclic Prefix}
\newacronym{up}{UP}{User Plane}
\newacronym{cpu}{CPU}{Central Processing Unit}
\newacronym{cqi}{CQI}{Channel Quality Information}
\newacronym{cr}{CR}{Cognitive Radio}
\newacronym{cnn}{CNN}{Convolutional Neural Network}
\newacronym{cran}{C-RAN}{Cloud \gls{ran}}
\newacronym{crs}{CRS}{Cell Reference Signal}
\newacronym{csi}{CSI}{Channel State Information}
\newacronym{csirs}{CSI-RS}{Channel State Information - Reference Signal}
\newacronym{cu}{CU}{Central Unit}
\newacronym{d2tcp}{D$^2$TCP}{Deadline-aware Data center TCP}
\newacronym{d3}{D$^3$}{Deadline-Driven Delivery}
\newacronym{dac}{DAC}{Digital to Analog Converter}
\newacronym{dag}{DAG}{Directed Acyclic Graph}
\newacronym{das}{DAS}{Distributed Antenna System}
\newacronym{dash}{DASH}{Dynamic Adaptive Streaming over HTTP}
\newacronym{dc}{DC}{Dual Connectivity}
\newacronym{dccp}{DCCP}{Datagram Congestion Control Protocol}
\newacronym{dce}{DCE}{Direct Code Execution}
\newacronym{dci}{DCI}{Downlink Control Information}
\newacronym{dctcp}{DCTCP}{Data Center TCP}
\newacronym{dl}{DL}{Downlink}
\newacronym{dmr}{DMR}{Deadline Miss Ratio}
\newacronym{dmrs}{DMRS}{DeModulation Reference Signal}
\newacronym{drlcc}{DRL-CC}{Deep Reinforcement Learning Congestion Control}
\newacronym{drs}{DRS}{Discovery Reference Signal}
\newacronym{du}{DU}{Distributed Unit}
\newacronym{e2e}{E2E}{end-to-end}
\newacronym{ecaas}{ECaaS}{Edge-Cloud-as-a-Service}
\newacronym{ecn}{ECN}{Explicit Congestion Notification}
\newacronym{edf}{EDF}{Earliest Deadline First}
\newacronym{embb}{eMBB}{Enhanced Mobile Broadband}
\newacronym{empower}{EMPOWER}{EMpowering transatlantic PlatfOrms for advanced WirEless Research}
\newacronym{enb}{eNB}{evolved Node Base}
\newacronym{endc}{EN-DC}{E-UTRAN-\gls{nr} \gls{dc}}
\newacronym{epc}{EPC}{Evolved Packet Core}
\newacronym{eps}{EPS}{Evolved Packet System}
\newacronym{es}{ES}{Edge Server}
\newacronym{etsi}{ETSI}{European Telecommunications Standards Institute}
\newacronym[firstplural=Estimated Times of Arrival (ETAs)]{eta}{ETA}{Estimated Time of Arrival}
\newacronym{eutran}{E-UTRAN}{Evolved Universal Terrestrial Access Network}
\newacronym{faas}{FaaS}{Function-as-a-Service}
\newacronym{fapi}{FAPI}{Functional Application Platform Interface}
\newacronym{fdd}{FDD}{Frequency Division Duplexing}
\newacronym{fdm}{FDM}{Frequency Division Multiplexing}
\newacronym{fdma}{FDMA}{Frequency Division Multiple Access}
\newacronym{fed4fire}{FED4FIRE+}{Federation 4 Future Internet Research and Experimentation Plus}
\newacronym{fir}{FIR}{Finite Impulse Response}
\newacronym{fit}{FIT}{Future \acrlong{iot}}
\newacronym{fpga}{FPGA}{Field Programmable Gate Array}
\newacronym{fr2}{FR2}{Frequency Range 2}
\newacronym{fs}{FS}{Fast Switching}
\newacronym{fscc}{FSCC}{Flow Sharing Congestion Control}
\newacronym{ftp}{FTP}{File Transfer Protocol}
\newacronym{fw}{FW}{Flow Window}
\newacronym{ge}{GE}{Gaussian Elimination}
\newacronym{gnb}{gNB}{Next Generation Node Base}
\newacronym{gop}{GOP}{Group of Pictures}
\newacronym{gpr}{GPR}{Gaussian Process Regressor}
\newacronym{gpu}{GPU}{Graphics Processing Unit}
\newacronym{gtp}{GTP}{GPRS Tunneling Protocol}
\newacronym{gtpc}{GTP-C}{GPRS Tunnelling Protocol Control Plane}
\newacronym{gtpu}{GTP-U}{GPRS Tunnelling Protocol User Plane}
\newacronym{gtpv2c}{GTPv2-C}{\gls{gtp} v2 - Control}
\newacronym{gw}{GW}{Gateway}
\newacronym{harq}{HARQ}{Hybrid Automatic Repeat reQuest}
\newacronym{hetnet}{HetNet}{Heterogeneous Network}
\newacronym{hh}{HH}{Hard Handover}
\newacronym{hol}{HOL}{Head-of-Line}
\newacronym{hqf}{HQF}{Highest-quality-first}
\newacronym{hss}{HSS}{Home Subscription Server}
\newacronym{http}{HTTP}{HyperText Transfer Protocol}
\newacronym{ia}{IA}{Initial Access}
\newacronym{iab}{IAB}{Integrated Access and Backhaul}
\newacronym{ic}{IC}{Incident Command}
\newacronym{isac}{ISAC}{integrated sensing and communication }
\newacronym{ietf}{IETF}{Internet Engineering Task Force}
\newacronym{imsi}{IMSI}{International Mobile Subscriber Identity}
\newacronym{imt}{IMT}{International Mobile Telecommunication}
\newacronym{iot}{IoT}{Internet of Things}
\newacronym{ip}{IP}{Internet Protocol}
\newacronym{itu}{ITU}{International Telecommunication Union}
\newacronym{jepa}{JEPA}{Joint Embedding Predictive Architecture}
\newacronym{kpi}{KPI}{Key Performance Indicator}
\newacronym{kpm}{KPM}{Key Performance Measurement}
\newacronym{kvm}{KVM}{Kernel-based Virtual Machine}
\newacronym{los}{LOS}{Line-of-Sight}
\newacronym{lsm}{LSM}{Link-to-System Mapping}
\newacronym{lstm}{LSTM}{Long Short Term Memory}
\newacronym{lte}{LTE}{Long Term Evolution}
\newacronym{lxc}{LXC}{Linux Container}
\newacronym{m2m}{M2M}{Machine to Machine}
\newacronym{mac}{MAC}{Medium Access Control}
\newacronym{manet}{MANET}{Mobile Ad Hoc Network}
\newacronym{mano}{MANO}{Management and Orchestration}
\newacronym{mc}{MC}{Multi-Connectivity}
\newacronym{mcc}{MCC}{Mobile Cloud Computing}
\newacronym{mchem}{MCHEM}{Massive Channel Emulator}
\newacronym{mcs}{MCS}{Modulation and Coding Scheme}
\newacronym{mec}{MEC}{Multi-access Edge Computing}
\newacronym{mec2}{MEC}{Mobile Edge Cloud}
\newacronym{mfc}{MFC}{Mobile Fog Computing}
\newacronym{mgen}{MGEN}{Multi-Generator}
\newacronym{mi}{MI}{Mutual Information}
\newacronym{mib}{MIB}{Master Information Block}
\newacronym{miesm}{MIESM}{Mutual Information Based Effective SINR}
\newacronym{mimo}{MIMO}{Multiple Input, Multiple Output}
\newacronym{ml}{ML}{Machine Learning}
\newacronym{mlr}{MLR}{Maximum-local-rate}
\newacronym[plural=\gls{mme}s,firstplural=Mobility Management Entities (MMEs)]{mme}{MME}{Mobility Management Entity}
\newacronym{mmtc}{mMTC}{Massive Machine-Type Communications}
\newacronym{mmwave}{mmWave}{millimeter wave}
\newacronym{mpdccp}{MP-DCCP}{Multipath Datagram Congestion Control Protocol}
\newacronym{mptcp}{MPTCP}{Multipath TCP}
\newacronym{mr}{MR}{Maximum Rate}
\newacronym{mrdc}{MR-DC}{Multi \gls{rat} \gls{dc}}
\newacronym{mse}{MSE}{Mean Square Error}
\newacronym{mss}{MSS}{Maximum Segment Size}
\newacronym{mt}{MT}{Mobile Termination}
\newacronym{mtd}{MTD}{Machine-Type Device}
\newacronym{mtu}{MTU}{Maximum Transmission Unit}
\newacronym{mumimo}{MU-MIMO}{Multi-user \gls{mimo}}
\newacronym{mvno}{MVNO}{Mobile Virtual Network Operator}
\newacronym{nalu}{NALU}{Network Abstraction Layer Unit}
\newacronym{nas}{NAS}{Non-Access Stratum}
\newacronym{nbiot}{NB-IoT}{Narrow Band IoT}
\newacronym{nfv}{NFV}{Network Function Virtualization}
\newacronym{nfvi}{NFVI}{Network Function Virtualization Infrastructure}
\newacronym{ngrg}{nGRG}{next Generation Research Group}
\newacronym{ni}{NI}{Network Interfaces}
\newacronym{nic}{NIC}{Network Interface Card}
\newacronym{nlos}{NLOS}{Non-Line-of-Sight}
\newacronym{now}{NOW}{Non Overlapping Window}
\newacronym{nsm}{NSM}{Network Service Mesh}
\newacronym{nr}{NR}{New Radio}
\newacronym{nrf}{NRF}{Network Repository Function}
\newacronym{nsa}{NSA}{Non Stand Alone}
\newacronym{nse}{NSE}{Network Slicing Engine}
\newacronym{nssf}{NSSF}{Network Slice Selection Function}
\newacronym{o2i}{O2I}{Outdoor to Indoor}
\newacronym{oai}{OAI}{OpenAirInterface}
\newacronym{oaicn}{OAI-CN}{\gls{oai} \acrlong{cn}}
\newacronym{oairan}{OAI-RAN}{\acrlong{oai} \acrlong{ran}}
\newacronym{oam}{OAM}{Operations, Administration and Maintenance}
\newacronym{ofdm}{OFDM}{Orthogonal Frequency Division Multiplexing}
\newacronym{olia}{OLIA}{Opportunistic Linked Increase Algorithm}
\newacronym{omec}{OMEC}{Open Mobile Evolved Core}
\newacronym{onap}{ONAP}{Open Network Automation Platform}
\newacronym{onf}{ONF}{Open Networking Foundation}
\newacronym{onos}{ONOS}{Open Networking Operating System}
\newacronym{oom}{OOM}{\gls{onap} Operations Manager}
\newacronym{opnfv}{OPNFV}{Open Platform for \gls{nfv}}
\newacronym{oran}{O-RAN}{Open Radio Access Network}
\newacronym{orbit}{ORBIT}{Open-Access Research Testbed for Next-Generation Wireless Networks}
\newacronym{os}{OS}{Operating System}
\newacronym{oss}{OSS}{Operations Support System}
\newacronym{otic}{OTIC}{Open Testing \& Integration Centre}
\newacronym{pa}{PA}{Position-aware}
\newacronym{pase}{PASE}{Prioritization, Arbitration, and Self-adjusting Endpoints}
\newacronym{pawr}{PAWR}{Platforms for Advanced Wireless Research}
\newacronym{pbch}{PBCH}{Physical Broadcast Channel}
\newacronym{pcef}{PCEF}{Policy and Charging Enforcement Function}
\newacronym{pcfich}{PCFICH}{Physical Control Format Indicator Channel}
\newacronym{pcrf}{PCRF}{Policy and Charging Rules Function}
\newacronym{pdcch}{PDCCH}{Physical Downlink Control Channel}
\newacronym{pdcp}{PDCP}{Packet Data Convergence Protocol}
\newacronym{pdsch}{PDSCH}{Physical Downlink Shared Channel}
\newacronym{pdu}{PDU}{Packet Data Unit}
\newacronym{pf}{PF}{Proportional Fair}
\newacronym{pgw}{PGW}{Packet Gateway}
\newacronym{phich}{PHICH}{Physical Hybrid ARQ Indicator Channel}
\newacronym{phy}{PHY}{Physical}
\newacronym{pmch}{PMCH}{Physical Multicast Channel}
\newacronym{pmi}{PMI}{Precoding Matrix Indicators}
\newacronym{powder}{POWDER}{Platform for Open Wireless Data-driven Experimental Research}
\newacronym{ppo}{PPO}{Proximal Policy Optimization}
\newacronym{ppp}{PPP}{Poisson Point Process}
\newacronym{prach}{PRACH}{Physical Random Access Channel}
\newacronym{prb}{PRB}{Physical Resource Block}
\newacronym{psnr}{PSNR}{Peak Signal to Noise Ratio}
\newacronym{pss}{PSS}{Primary Synchronization Signal}
\newacronym{pucch}{PUCCH}{Physical Uplink Control Channel}
\newacronym{pusch}{PUSCH}{Physical Uplink Shared Channel}
\newacronym{rar}{RAR}{Random Access Response}
\newacronym{qam}{QAM}{Quadrature Amplitude Modulation}
\newacronym{qci}{QCI}{\gls{qos} Class Identifier}
\newacronym{5qi}{5QI}{5G \gls{qos} Identifier}
\newacronym{qoe}{QoE}{Quality of Experience}
\newacronym{QoS}{QoS}{Quality of Service}
\newacronym{UE}{UE}{User Equipment}
\newacronym{UEs}{UEs}{User Equipments}
\newacronym{FoV}{FoV}{field of view}
\newacronym{UPA}{UPA}{uniform planar array}
\newacronym{quic}{QUIC}{Quick UDP Internet Connections}
\newacronym{rach}{RACH}{Random Access Channel}
\newacronym{ran}{RAN}{Radio Access Network}
\newacronym[firstplural=Radio Access Technologies (RATs)]{rat}{RAT}{Radio Access Technology}
\newacronym{rcn}{RCN}{Research Coordination Network}
\newacronym{STAR}{STAR-RIS}{simultaneous transmitting and reflecting RIS}
\newacronym{3SNCR}{3SNCR}{trisectoral NCR}
\newacronym{rc}{RC}{RAN Control}
\newacronym{rec}{REC}{Radio Edge Cloud}
\newacronym{red}{RED}{Random Early Detection}
\newacronym{renew}{RENEW}{Reconfigurable Eco-system for Next-generation End-to-end Wireless}
\newacronym{rf}{RF}{Radio Frequency}
\newacronym{rfc}{RFC}{Request for Comments}
\newacronym{rfr}{RFR}{Random Forest Regressor}
\newacronym{ric}{RIC}{\gls{ran} Intelligent Controller}
\newacronym{rlc}{RLC}{Radio Link Control}
\newacronym{rlf}{RLF}{Radio Link Failure}
\newacronym{rlnc}{RLNC}{Random Linear Network Coding}
\newacronym{rmr}{RMR}{RIC Message Router}
\newacronym{rmse}{RMSE}{Root Mean Squared Error}
\newacronym{rnis}{RNIS}{Radio Network Information Service}
\newacronym{rr}{RR}{Round Robin}
\newacronym{rrc}{RRC}{Radio Resource Control}
\newacronym{rrm}{RRM}{Radio Resource Management}
\newacronym{rru}{RRU}{Remote Radio Unit}
\newacronym{rs}{RS}{Remote Server}
\newacronym{rsrp}{RSRP}{Reference Signal Received Power}
\newacronym{rsrq}{RSRQ}{Reference Signal Received Quality}
\newacronym{rss}{RSS}{Received Signal Strength}
\newacronym{rssi}{RSSI}{Received Signal Strength Indicator}
\newacronym{rtt}{RTT}{Round Trip Time}
\newacronym{ru}{RU}{Radio Unit}
\newacronym{rw}{RW}{Receive Window}
\newacronym{rx}{RX}{Receiver}
\newacronym{rnn}{RNN}{Recurrent Neural Network}
\newacronym{s1ap}{S1AP}{S1 Application Protocol}
\newacronym{sa}{SA}{standalone}
\newacronym{sack}{SACK}{Selective Acknowledgment}
\newacronym{sap}{SAP}{Service Access Point}
\newacronym{sc2}{SC2}{Spectrum Collaboration Challenge}
\newacronym{scef}{SCEF}{Service Capability Exposure Function}
\newacronym{sch}{SCH}{Secondary Cell Handover}
\newacronym{scoot}{SCOOT}{Split Cycle Offset Optimization Technique}
\newacronym{sctp}{SCTP}{Stream Control Transmission Protocol}
\newacronym{sdap}{SDAP}{Service Data Adaptation Protocol}
\newacronym{sdk}{SDK}{Software Development Kit}
\newacronym{sdm}{SDM}{Space Division Multiplexing}
\newacronym{sdma}{SDMA}{Spatial Division Multiple Access}
\newacronym{sdn}{SDN}{Software-defined Networking}
\newacronym{sdr}{SDR}{Software-defined Radio}
\newacronym{seba}{SEBA}{SDN-Enabled Broadband Access}
\newacronym{sgsn}{SGSN}{Serving GPRS Support Node}
\newacronym{sgw}{SGW}{Service Gateway}
\newacronym{si}{SI}{Study Item}
\newacronym{sib}{SIB}{Secondary Information Block}
\newacronym{sinr}{SINR}{Signal to Interference plus Noise Ratio}
\newacronym{sip}{SIP}{Session Initiation Protocol}
\newacronym{siso}{SISO}{Single Input, Single Output}
\newacronym{sla}{SLA}{Service Level Agreement}
\newacronym{sm}{SM}{Service Model}
\newacronym{mae}{MAE}{masked autoencoder}
\newacronym{smo}{SMO}{Service Management and Orchestration}
\newacronym{sms}{SMS}{Short Message Service}
\newacronym{smsgmsc}{SMS-GMSC}{\gls{sms}-Gateway}
\newacronym{snr}{SNR}{Signal-to-Noise-Ratio}
\newacronym{son}{SON}{Self-Organizing Network}
\newacronym{sptcp}{SPTCP}{Single Path TCP}
\newacronym{srb}{SRB}{Service Radio Bearer}
\newacronym{srn}{SRN}{Standard Radio Node}
\newacronym{srs}{SRS}{Sounding Reference Signal}
\newacronym{zc}{ZC}{Zadoff-Chu}
\newacronym{ta}{TA}{Timing Advance}
\newacronym{ss}{SS}{Synchronization Signal}
\newacronym{ssl}{SSL}{Self-Supervised Learning}
\newacronym{sss}{SSS}{Secondary Synchronization Signal}
\newacronym{st}{ST}{Spanning Tree}
\newacronym{svc}{SVC}{Scalable Video Coding}
\newacronym{tb}{TB}{Transport Block}
\newacronym{tcp}{TCP}{Transmission Control Protocol}
\newacronym{tdd}{TDD}{Time Division Duplexing}
\newacronym{tdm}{TDM}{Time Division Multiplexing}
\newacronym{tdma}{TDMA}{Time Division Multiple Access}
\newacronym{tfl}{TfL}{Transport for London}
\newacronym{tfrc}{TFRC}{TCP-Friendly Rate Control}
\newacronym{tft}{TFT}{Traffic Flow Template}
\newacronym{tgen}{TGEN}{Traffic Generator}
\newacronym{tip}{TIP}{Telecom Infra Project}
\newacronym{tm}{TM}{Transparent Mode}
\newacronym{to}{TO}{Telco Operator}
\newacronym{tr}{TR}{Technical Report}
\newacronym{trp}{TRP}{Transmitter Receiver Pair}
\newacronym{ts}{TS}{Technical Specification}
\newacronym{tti}{TTI}{Transmission Time Interval}
\newacronym{ttt}{TTT}{Time-to-Trigger}
\newacronym{tx}{TX}{Transmitter}
\newacronym{uas}{UAS}{Unmanned Aerial System}
\newacronym{uav}{UAV}{Unmanned Aerial Vehicle}
\newacronym{udm}{UDM}{Unified Data Management}
\newacronym{udp}{UDP}{User Datagram Protocol}
\newacronym{udr}{UDR}{Unified Data Repository}
\newacronym{ue}{UE}{User Equipment}
\newacronym{uhd}{UHD}{\gls{usrp} Hardware Driver}
\newacronym{ul}{UL}{Uplink}
\newacronym{um}{UM}{Unacknowledged Mode}
\newacronym{uml}{UML}{Unified Modeling Language}
\newacronym{upa}{UPA}{Uniform Planar Array}
\newacronym{upf}{UPF}{User Plane Function}
\newacronym{urllc}{URLLC}{Ultra Reliable and Low Latency Communications}
\newacronym{usa}{U.S.}{United States}
\newacronym{usim}{USIM}{Universal Subscriber Identity Module}
\newacronym{usrp}{USRP}{Universal Software Radio Peripheral}
\newacronym{utc}{UTC}{Urban Traffic Control}
\newacronym{vim}{VIM}{Virtualization Infrastructure Manager}
\newacronym{vm}{VM}{Virtual Machine}
\newacronym{vnf}{VNF}{Virtual Network Function}
\newacronym{volte}{VoLTE}{Voice over \gls{lte}}
\newacronym{voltha}{VOLTHA}{Virtual OLT HArdware Abstraction}
\newacronym{vr}{VR}{Virtual Reality}
\newacronym{vran}{vRAN}{Virtualized \gls{ran}}
\newacronym{vss}{VSS}{Video Streaming Server}
\newacronym{wbf}{WBF}{Wired Bias Function}
\newacronym{wf}{WF}{Waterfilling}
\newacronym{wg}{WG}{Working Group}
\newacronym{wlan}{WLAN}{Wireless Local Area Network}
\newacronym{osm}{OSM}{Open Source Management and Orchestration}
\newacronym{pnf}{PNF}{Physical Network Function}
\newacronym{drl}{DRL}{Deep Reinforcement Learning}
\newacronym{mtc}{MTC}{Machine-type Communications}
\newacronym{osc}{OSC}{O-RAN Software Community}
\newacronym{mns}{MnS}{Management Services}
\newacronym{ves}{VES}{\gls{vnf} Event Stream}
\newacronym{ei}{EI}{Enrichment Information}
\newacronym{fh}{FH}{Fronthaul}
\newacronym{fft}{FFT}{Fast Fourier Transform}
\newacronym{laa}{LAA}{Licensed-Assisted Access}
\newacronym{plfs}{PLFS}{Physical Layer Frequency Signals}
\newacronym{ptp}{PTP}{Precision Time Protocol}
\newacronym{asic}{ASIC}{Application-specific Integrated Circuit}
\newacronym{aal}{AAL}{Acceleration Abstraction Layer}
\newacronym{fec}{FEC}{Forward Error Correction}
\newacronym{sdl}{SDL}{Shared Data Layer}
\newacronym{nib}{NIB}{Network Information Base}
\newacronym{rnib}{R-NIB}{RAN \gls{nib}}
\newacronym{fcaps}{FCAPS}{Fault, Configuration, Accounting, Performance, Security}
\newacronym{ie}{IE}{Information Element}
\newacronym{fg}{FG}{Focus Group}
\newacronym{osfg}{OSFG}{Open Source Focus Group}
\newacronym{sdfg}{SDFG}{Standard Development Focus Group}
\newacronym{tifg}{TIFG}{Test \& Integration Focus Group}
\newacronym{sfg}{SFG}{Security Focus Group}
\newacronym{swg}{SWG}{Security Work Group}
\newacronym{e2sm}{E2SM}{E2 Service Model}
\newacronym{tsc}{TSC}{Technical Steering Committee}
\newacronym{sdo}{SDO}{Standard-Development Organization}
\newacronym{sql}{SQL}{Structured Query Language}
\newacronym{ssh}{SSH}{Secure Shell}
\newacronym{tls}{TLS}{Transport Layer Security}
\newacronym{netconf}{NETCONF}{Network Configuration Protocol}
\newacronym{dtls}{DTLS}{Datagram Transport Layer Security}
\newacronym{cmp}{CMP}{Certificate Management Protocol}
\newacronym{ccc}{CCC}{Cell Configuration and Control}
\newacronym{dsp}{DSP}{Digital Signal Processing}
\newacronym{opex}{OPEX}{Operational Expenses}
\newacronym{cbrs}{CBRS}{Citizen Broadband Radio Service}
\newacronym{ntn}{NTN}{Non-terrestrial Network}
\newacronym{gbr}{GBR}{Guaranteed Bitrate}
\newacronym{sps}{SPS}{Semi-Persistent Scheduling}
\newacronym{tbs}{TBS}{Transport Block Size}
\newacronym{gnss}{GNSS}{Global Navigation Satellite System}
\newacronym{tof}{ToF}{Time of Flight}
\newacronym{rtof}{RToF}{Return Time of Flight}
\newacronym{rsig}{RS}{Reference Signal}
\newacronym{nrtric}{near-RT RIC}{near-Real Time Ran Intelligent Controller}
\newacronym{nonrtric}{non-RT RIC}{non-Real Time Ran Intelligent Controller}
\newacronym{aoa}{AoA}{Angle of Arrival}
\newacronym{tdoa}{TDoA}{Time Difference of Arrival}
\newacronym{rtoa}{RToA}{Return Time of Arrival}
\newacronym{ris}{RIS}{Reconfigurable Intelligent Surface}
\newacronym{srd}{SRD}{Smart Radio Device}
\newacronym{gfbr}{GFBR}{Guaranteed Flow Bit Rate}
\newacronym{rg}{RG}{Resource Grid}
\newacronym{rb}{RB}{Resource Block}
\newacronym{re}{RE}{Resource Element}
\newacronym{rfra}{RF}{Radio Frame}
\newacronym{scs}{SCS}{Subcarrier Spacing}
\newacronym{ec}{EC}{Edge Computing}
\newacronym{af}{AF}{Amplify-and-Forward}
\newacronym{ncr}{NCR}{Network-Controlled Repeater}
\newacronym{tp}{TP}{Test Point}
\newacronym{cs}{CS}{Candidate Site}
\newacronym{src}{SRC}{Smart Radio Connection}
\newacronym{milp}{MILP}{Mixed Integer-Linear Programming}

\newacronym{FCMC}{FCMC}{full coverage minimum cost}

\newacronym{MBCC}{MBCC}{maximum budget-constrained coverage}

\newacronym{PDF}{PDF}{probability density function}
\newacronym{tsne}{t-SNE}{t-distributed stochastic neighbor 
embedding}
\newacronym{stsne}{St-SNE}{Semi-Supervised t-distributed stochastic neighbor 
embedding}
\newacronym{vit}{ViT}{Vision Transformer}
\newacronym{lwlm}{LWLM}{Lightweight Localization Module}

\begin{document}

\title{Masked Latent Prediction of CSI for Indoor Localization in Integrated Sensing and Communication Systems}

\author{
\IEEEauthorblockN{
Ibtissam Labriji\IEEEauthorrefmark{1}\textsuperscript{\(\ddagger\)},
Mahdi Maleki\IEEEauthorrefmark{2}\textsuperscript{\(\ddagger\)},
and Pavan Koteshwar Srinath\IEEEauthorrefmark{1}
}
\IEEEauthorblockA{\IEEEauthorrefmark{1}
Nokia Bell Labs, France\\
Email: \{ibtissam.labriji, pavan.koteshwar\_srinath\}@nokia-bell-labs.com
}
\IEEEauthorblockA{\IEEEauthorrefmark{2}
Politecnico di Milano, Italy\\
Email: mahdi.maleki@polimi.it
}
\IEEEauthorblockA{
\footnotesize \(\ddagger\)These authors contributed equally to this work.
}
\thanks{This work has been submitted to the IEEE for possible
publication. Copyright may be transferred without notice,
after which this version may no longer be accessible.}
}

\maketitle

\begin{abstract}
\gls{csi}-based fingerprinting can enable accurate indoor localization but
suffers from domain shift, limited labeled data, and degraded performance in
multipath-rich environments. To address these challenges, we propose a
self-supervised localization framework built on a \gls{jepa}. Each \gls{csi}
time snapshot is treated as a token, and the encoder is pre-trained by
predicting the latent embeddings of masked snapshots from the visible ones,
learning robust channel representations without any labels. The encoder is
then frozen, and only a lightweight regression head is trained on a small set
of labeled positions to estimate the user location. We evaluate the framework
on the measured DICHASUS-005$x$ dataset---a single-antenna transmitter
received by a $32$-antenna array in a multipath-rich indoor environment.
With a $50\%$ masking ratio, the proposed method reduces the mean localization
error from $0.90$\,m to $0.42$\,m (a $53\%$ reduction, or $0.48$\,m) relative
to a supervised \gls{cnn} baseline trained on raw \gls{csi}. Owing to its
label-efficient, frozen-encoder design, the framework aligns with \gls{isac}
objectives, enabling reliable sensing with minimal labeling and compute
overhead. 

\end{abstract}


\glsresetall


\section{Introduction}

Fingerprinting exploits location-specific characteristics of radio
signals---\gls{rss}, \gls{csi}, and hardware-induced impairments---to localize a
device without dedicated infrastructure or transmitter cooperation
\cite{fingerPrint_general}. Among these, \gls{csi}-based fingerprinting is
especially attractive indoors due to its fine-grained view of the propagation
environment, and deep models (\glspl{cnn}, \glspl{rnn}, and Transformers) learn
discriminative features from high-dimensional \gls{csi} in dense settings where
\gls{GPS} is unreliable \cite{ssl_survey,stahlke2023indoor}. Such localization
is also central to \gls{isac}, where communication signals are reused to
estimate spatial parameters such as position, angle, and velocity without
sensing-dedicated hardware---a key direction for future 6G networks
\cite{isac_survey}. Fingerprints, however, are highly sensitive to domain
shift: environmental changes, mobility, and varying signal conditions degrade
accuracy relative to the training-time radio map \cite{chartwin}.
A further obstacle is the reliance on large, accurately labeled radio maps,
whose collection and periodic refreshing is costly in both semi-supervised
\cite{maleki2025towards,chartwin} and fully-supervised
\cite{fingerPrint_general} settings, and is compounded by severe multipath,
blockage, and frequent \gls{nlos} conditions \cite{JEPA_CC}. While supervised
models can capture complex \gls{csi} correlations and improve robustness under
\gls{nlos}, their accuracy still drops sharply under environmental change
\cite{CC_UAV}.
To reduce this label dependence, \gls{ssl} learns transferable representations
directly from unlabeled signals \cite{ssl_survey}, via masked modeling,
contrastive, or predictive objectives \cite{jepa_tut}. Yet contrastive methods
require hand-crafted, physics-preserving augmentations that are difficult to
design for \gls{rf} signals, while reconstruction-based methods expend capacity
on low-level detail largely irrelevant to position. The \gls{jepa} avoids both
by predicting abstract latent representations rather than reconstructing raw
observations \cite{monemi2025tutorial}, concentrating capacity on semantically
meaningful, geometry-linked structure. Recent WiFi-sensing work shows that
channel information can indeed be predicted in latent space through cross-modal
semantic inpainting \cite{nishio2025semantic}. Combined with the label scarcity
and the difficulty of designing physics-preserving \gls{rf} augmentations noted
above, this makes \gls{jepa}'s augmentation-free latent prediction a natural fit
for \gls{csi}-based localization; yet such prediction \emph{within} the
intrinsic structure of \gls{csi}---and the tokenization and masking design that
yields label-efficient, environment-invariant representations robust to
multipath, \gls{nlos}, and missing measurements---remains largely unexplored
\cite{maleki2025channel}.

\subsection{Motivation and Contributions}

Motivated by these challenges, this paper investigates indoor localization
using \gls{jepa}-based representation learning. We consider an indoor region of
interest in which a single-antenna user transmits to a multi-antenna \gls{bs},
and localization is performed from \gls{csi} fingerprints extracted from the
received signal. The key idea is to learn, through masked temporal prediction
in latent space, an embedding in which observations from nearby positions
become more structurally informative and easier to localize, before any labels
are used. The main contributions of this paper are:
\begin{itemize}
  \item We propose a \gls{jepa}-based self-supervised framework for \gls{csi}
        localization that treats each time snapshot as a token and predicts the
        latent representations of masked snapshots from the visible ones,
        avoiding both augmentation design and raw-\gls{csi} reconstruction.
  \item We adopt a two-stage design in which the encoder is pre-trained once
        and \emph{frozen}, and only a lightweight regression head is trained on
        a small set of labeled positions, enabling label-efficient and
        low-latency deployment at the network edge.
  \item We validate the approach on the measured DICHASUS dataset (a
        single-antenna transmitter and a $32$-antenna receiver) and show that
        it lowers the mean localization error by a factor of about $2.1\times$
        (from $0.90$\,m to $0.42$\,m) relative to a supervised \gls{cnn}
        trained on raw \gls{csi}, with a consistently left-shifted error CDF.
\end{itemize}

The remainder of this paper is organized as follows.
Section~\ref{sec:sota} reviews related work on indoor localization,
fingerprinting, and self-supervised representation learning.
Section~\ref{sec:model} presents the system model and fingerprint
construction. Section~\ref{sec:method} introduces the proposed \gls{jepa}-based
framework. Section~\ref{sec:eval} reports the experimental results, and
Section~\ref{sec:conclusion} concludes the paper.

\section{State of the Art}
\label{sec:sota}

\gls{csi} fingerprinting with \gls{cnn}, \gls{rnn}, and Transformer
architectures has achieved high accuracy in dense indoor deployments by
learning discriminative features over subcarriers and antennas
\cite{fingerPrint_general,ssl_survey}. These methods, however, are typically
trained per environment and degrade under domain shift induced by temporal
variation, furniture rearrangement, changes in access-point configuration,
and human presence. Supervised pipelines further require large, precisely
labeled radio maps, whose collection and periodic refreshing is costly in
complex indoor spaces. Semi-supervised and domain-adaptation variants reduce
the labeling burden but still struggle to maintain stability across sites and
over time. A central difficulty is that multipath propagation and
\gls{nlos} conditions \cite{chartwin} lead these models to learn entangled
representations that cannot separate the permanent room geometry from
transient changes such as moving people, so even attention-augmented ResNets
and \glspl{vit} lose accuracy when the environment changes slightly
\cite{ssl_survey}.

Self-supervised learning has been explored to reduce label dependence, but
the dominant paradigms transfer poorly to \gls{rf} data. Contrastive methods
rely on hand-crafted augmentations, and determining which transformations
preserve the underlying channel physics while remaining informative requires
substantial expert tuning \cite{jepa_tut}. Reconstruction-based
approaches such as \gls{mae} instead expend model capacity on regenerating
low-level \gls{csi} detail that is largely irrelevant to localization. In
contrast, predictive latent objectives---exemplified by \gls{jepa}---learn by
predicting masked context in representation space, avoiding both augmentation
design and pixel-level reconstruction, and thereby concentrating capacity on
stable, geometry-linked factors of variation
\cite{monemi2025tutorial,JEPA_CC}.

Beyond \gls{csi} fingerprinting, several ranging-based technologies provide
indoor positioning under specific hardware and protocol assumptions.
Wi-Fi Round-Trip Time (802.11mc/az) estimates distance from signal
time-of-flight and typically achieves $1$--$2$\,m accuracy when access points support
\gls{FTM}, degrading to $3$--$4$\,m for one-sided RTT with legacy hardware
\cite{nishio2025semantic}. \gls{BLE}~5.1 Angle-of-Arrival/Departure (AoA/AoD) can reach sub-meter
accuracy with multi-antenna anchors---an empirical study reports an average
error of about $0.7$\,m---but remains sensitive to multipath, antenna-array
calibration, and anchor placement \cite{pau2021bluetooth}. \gls{UWB} offers the
highest ranging precision at $10$--$30$\,cm under clear \gls{los}, at the cost
of dedicated hardware and tight timing synchronization \cite{UWB_loc}. These
technologies deliver strong accuracy only under favorable conditions and
typically depend on specialized hardware, calibration, or protocol support,
in contrast to \gls{csi}-based approaches that reuse existing
Wi-Fi/cellular infrastructure.

In summary, prior art either (i) attains high accuracy but is fragile under
domain shift and label-hungry, (ii) applies self-supervision in forms poorly
matched to \gls{rf} physics, or (iii) requires dedicated ranging hardware.
This motivates a label-efficient, augmentation-free representation that is
robust to multipath and \gls{nlos} and deployable on existing infrastructure,
which we pursue through masked temporal latent prediction over \gls{csi}.

\section{System Model}
\label{sec:model}

\begin{figure*}[t]
    \centering
    \includegraphics[
        width=0.95\textwidth,
        trim=0cm 6cm 0cm 6cm,
        clip
    ]{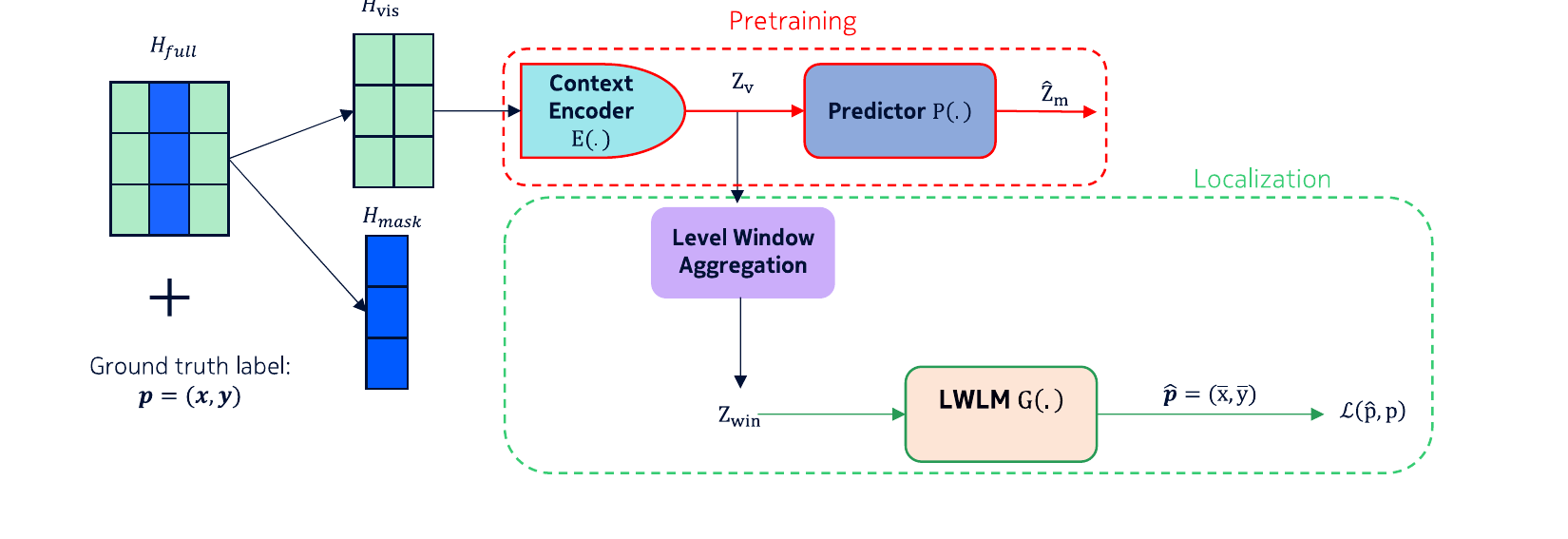}
    \caption{Proposed localization pipeline. A \gls{csi} window
    $H_{\text{full}}$ is split into visible ($H_{\text{vis}}$) and masked
    ($H_{\text{mask}}$) snapshots. During pre-training, $E(\cdot)$ produces
    $Z_v$ and $P(\cdot)$ predicts $\hat Z_m$; for localization, the frozen
    encoder's latents are aggregated into $Z_{\text{win}}$, which the
    \gls{lwlm} $G(\cdot)$ maps to $\hat{\mathbf p}$, supervised by the
    ground-truth label $\mathbf p=(x,y)$.}
    \label{fig:system}
\end{figure*}

We consider an indoor wireless localization system that exploits \gls{csi}
measurements collected from a radio infrastructure such as a \gls{bs} or
\gls{ap}. The objective is to estimate the spatial position of a \gls{ue} from
high-dimensional \gls{csi} observations under realistic propagation conditions,
including multipath, \gls{nlos}, and irregular temporal sampling. The proposed
framework follows a masked temporal latent prediction pipeline, illustrated in
Fig.~\ref{fig:system}.

\subsection{Overview}
In practice, a \gls{bs} or \gls{ap} continuously logs \gls{csi} as users move
through the area, yielding large amounts of unlabeled measurements but very few
position-annotated ones. We exploit this asymmetry with two stages that share a
single encoder (Fig.~\ref{fig:system}). Offline, the encoder is pre-trained on
the unlabeled logs alone: each window is split into visible and masked
snapshots, and the encoder learns to predict the masked snapshots in latent
space, supervised only by an EMA copy of itself (Fig.~\ref{fig:jepa}). This
stage needs no labels and can run on historical data. The encoder is then frozen
and, using a small set of surveyed reference points, only a lightweight head is
trained to map its embeddings to coordinates---a step cheap enough to repeat
when a site changes. At run time, localizing a user reduces to one frozen-encoder
pass over the visible snapshots followed by the small head, averaged over a few
masking patterns for stability, so inference stays fast and edge-friendly.

\subsection{\gls{csi} Signal Representation}
Let the \gls{csi} measurement at time index $t$ be denoted as
\begin{equation}
\mathbf{X}_t \in \mathbb{C}^{N_f \times N_a},
\end{equation}
where $N_f$ is the number of frequency bins and $N_a$ is the number of receive
antenna elements. A temporal window of \gls{csi} measurements is
\begin{equation}
H_{\text{full}} = \{ \mathbf{X}_1, \mathbf{X}_2, \dots, \mathbf{X}_T \}.
\end{equation}
Due to practical system constraints, the \gls{csi} observations within a window
may be irregularly sampled or partially missing.

\subsection{Temporal Tokenization}
Each \gls{csi} snapshot $\mathbf{X}_t$ is treated as a single time-aligned token
\begin{equation}
\mathcal{T} = \{ \tau_1, \tau_2, \dots, \tau_T \}, \qquad \tau_t = \mathbf{X}_t,
\end{equation}
where each token spans the full frequency--antenna structure. This formulation
preserves the temporal evolution of the propagation environment and avoids
splitting the \gls{csi} along the frequency or antenna dimensions.

\section{Proposed Method}
\label{sec:method}

\begin{figure*}[t]
    \centering
    \includegraphics[
        width=0.9\textwidth,
        trim=0cm 6cm 0cm 6cm,
        clip
    ]{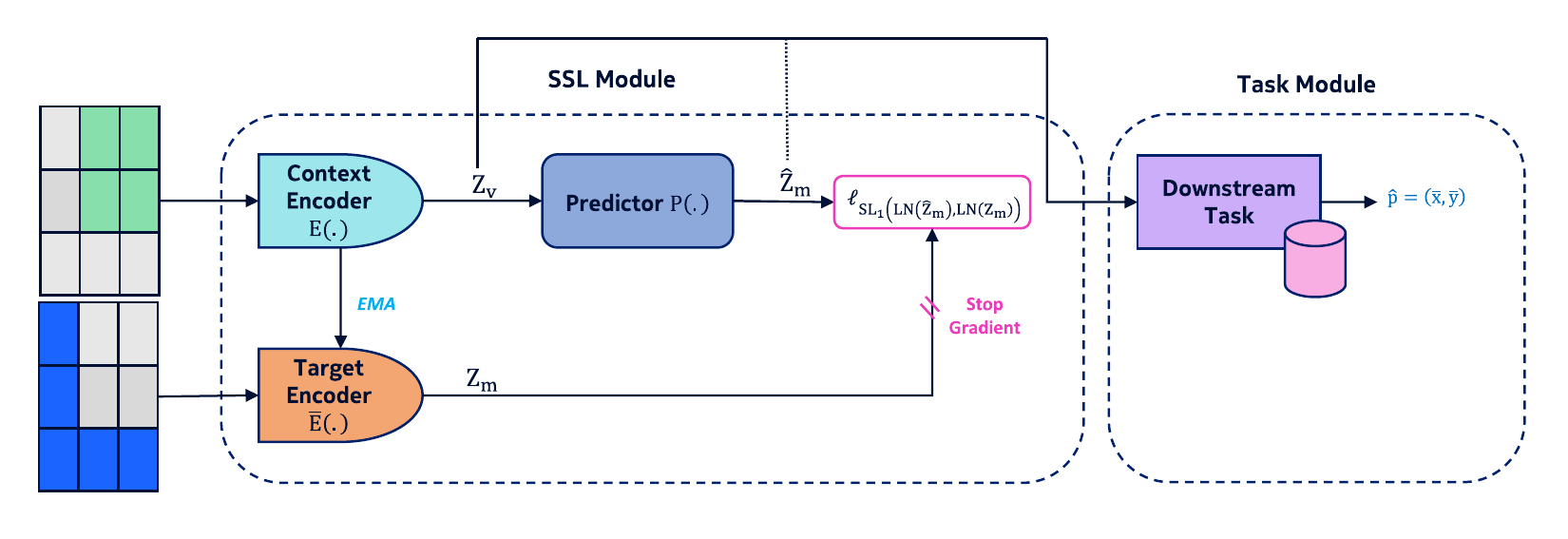}
    \caption{Self-supervised JEPA principle. The context encoder $E(\cdot)$
    encodes the visible snapshots $H_{\text{vis}}$ into latents $Z_v$, and the
    predictor $P(\cdot)$ estimates the latents $\hat Z_m$ of the masked
    snapshots $H_{\text{mask}}$. Targets $Z_m$ come from the EMA target encoder
    $\bar E(\cdot)$ under stop-gradient; training minimizes
    $\ell_{\text{SL1}}(\mathrm{LN}(\hat Z_m),\mathrm{LN}(Z_m))$. After
    pre-training, the frozen encoder feeds the \gls{lwlm} $G(\cdot)$.}
    \label{fig:jepa}
\end{figure*}

We present the proposed masked temporal latent prediction framework for
\gls{csi}-based localization. The underlying self-supervised principle is
summarized in Fig.~\ref{fig:jepa}, and its instantiation in the full
localization pipeline is shown in Fig.~\ref{fig:system}.

\subsection{Masked Temporal Learning}
The token set of a window is partitioned into a visible subset $H_{\text{vis}}$
and a masked subset $H_{\text{mask}}$, indexed respectively by the time-index
sets $\mathcal{V}$ and $\mathcal{M}$:
\begin{equation}
H_{\text{full}} = H_{\text{vis}} \cup H_{\text{mask}}, \qquad
\mathcal{V} \cap \mathcal{M} = \emptyset,\ \ \mathcal{V} \cup \mathcal{M} = \{1,\dots,T\}.
\end{equation}
Masking is applied to \emph{entire} \gls{csi} snapshots rather than partial
features, so each masked unit is a complete measurement occasion. During
pre-training the partition is drawn at random with a fixed masking ratio; for
evaluation it is generated from deterministic, coverage-balanced patterns
(Section~\ref{sec:eval}). Because each masked unit is a complete snapshot, the encoder must infer it from temporal context alone, which by construction makes it tolerant to missing or
irregularly sampled measurements and pushes it toward geometry-aware structure.

\subsection{Encoder--Predictor Architecture}
A context (online) encoder $E(\cdot)$ maps the visible tokens to latent
embeddings,
\begin{equation}
\mathbf{Z}_v = E(\mathbf{X}_v), \qquad v \in \mathcal{V},
\end{equation}
and a predictor $P(\cdot)$ estimates the latent representation of each masked
token from the visible embeddings and the masked position index,
\begin{equation}
\hat{\mathbf{Z}}_m = P\!\left(\{\mathbf{Z}_v\}_{v\in\mathcal{V}},\, m\right),
\qquad m \in \mathcal{M}.
\end{equation}
The prediction targets are produced by a target encoder $\bar{E}(\cdot)$, an
exponential moving average (EMA) of the online encoder that is not updated by
gradient descent,
\begin{equation}
\mathbf{Z}_m = \operatorname{sg}\!\big[\bar{E}(\mathbf{X}_m)\big],
\qquad
\bar{E} \leftarrow \rho\,\bar{E} + (1-\rho)\,E,
\end{equation}
where $\operatorname{sg}[\cdot]$ is the stop-gradient operator and
$\rho \in [0,1)$ is the EMA momentum (increased from $0.96$ to $0.99$ during
training). The EMA target together with the stop-gradient prevent representation collapse. This is what lets the objective remain augmentation-free: with slowly-moving, stop-gradient targets the predictor cannot collapse the representation, so no negatives or hand-crafted \gls{rf}
augmentations are required.

\subsection{Latent Prediction Objective}
Training minimizes a latent-space prediction loss between the predicted and
target embeddings of the masked tokens,
\begin{equation}
\mathcal{L} = \sum_{m \in \mathcal{M}}
\ell_{\text{SL1}}\!\Big(\operatorname{LN}(\hat{\mathbf{Z}}_m),\,
\operatorname{LN}(\mathbf{Z}_m)\Big),
\label{eq:loss}
\end{equation}
where $\operatorname{LN}(\cdot)$ denotes layer normalization applied over the
embedding dimension,
\begin{equation}
\operatorname{LN}(\mathbf{Z}) = \frac{\mathbf{Z}-\mu}{\sqrt{\sigma^2+\epsilon}},
\quad
\mu = \frac{1}{D}\sum_{d=1}^{D} Z_d,
\quad
\sigma^2 = \frac{1}{D}\sum_{d=1}^{D}(Z_d-\mu)^2,
\end{equation}
with $D$ the encoder embedding dimension and $\epsilon$ a small constant for
numerical stability. Layer normalization standardizes each $D$-dimensional
latent to zero mean and unit variance, so the comparison is invariant to scale
and offset and the targets remain stable during training. The term
$\ell_{\text{SL1}}$ is the smooth-$L_1$ (Huber) loss with threshold $\beta$,
\begin{equation}
\ell_{\text{SL1}}(r) =
\begin{cases}
\dfrac{r^2}{2\beta}, & |r| < \beta,\\[1mm]
|r| - \tfrac{\beta}{2}, & |r| \ge \beta,
\end{cases}
\end{equation}
applied element-wise and averaged over the embedding dimension, with
$\beta=1$. It behaves quadratically for small residuals, ensuring smooth
gradients, and linearly for large residuals, limiting the influence of outlier
dimensions caused by multipath and \gls{nlos} fluctuations. Operating purely in
latent space, this objective avoids modeling low-level \gls{csi} detail and
instead captures temporally consistent channel structure.

\subsection{Lightweight Localization Module}
At localization time the predictor $P(\cdot)$ is discarded, and the window
representation is built solely from the frozen encoder's embeddings of the
visible tokens, concatenated in temporal order,
\begin{equation}
\mathbf{Z}_{\text{win}} = [\mathbf{Z}_1 \parallel \mathbf{Z}_2 \parallel \dots \parallel \mathbf{Z}_K],
\qquad K = |\mathcal{V}|,
\end{equation}
where $\parallel$ denotes concatenation. With a fixed masking ratio, $K$ is
constant, so $\mathbf{Z}_{\text{win}}$ has a fixed dimension. A \gls{lwlm}
$G(\cdot)$ then maps this representation to a position estimate,
\begin{equation}
\hat{\mathbf{p}} = G(\mathbf{Z}_{\text{win}}),
\qquad \hat{\mathbf{p}} \in \mathbb{R}^2\ \text{or}\ \mathbb{R}^3.
\end{equation}
The \gls{lwlm} is a small multi-layer perceptron and is the only component
trained with labeled data; the encoder $E(\cdot)$ remains frozen. At inference,
the estimate is averaged over $L$ deterministic balanced masks
$\{\mathcal{V}^{(j)}\}_{j=1}^{L}$ to obtain a mask-invariant position,
\begin{equation}
\hat{\mathbf{p}} = (\bar{x},\bar{y}) = \frac{1}{L}\sum_{j=1}^{L}
G\!\big(\mathbf{Z}_{\text{win}}^{(j)}\big),
\end{equation}
which is compared against the ground-truth position $\mathbf{p}=(x,y)$.

\subsection{Training Strategy}
The framework follows a two-stage procedure:
\begin{itemize}
  \item \textbf{Representation learning (offline, unlabeled):} the online
        encoder $E$ and predictor $P$ are trained by masked temporal latent
        prediction, with the target encoder $\bar{E}$ updated by EMA.
  \item \textbf{Localization (labeled):} the encoder is frozen, and only the
        \gls{lwlm} $G$ is trained on labeled positions.
\end{itemize}
Decoupling representation learning from localization reduces labeling
requirements and online latency by design, and is expected to ease cross-site
adaptation, since only the small module $G$ must be retrained for a new
environment.

\section{Performance Evaluation}
\label{sec:eval}

We evaluate on the DICHASUS-005x measured channel dataset~\cite{dichasus_sounder,dichasus_dataset},
in which a single-antenna transmitter mounted on a mobile robot is received by
a $32$-antenna array (element spacing $0.118$\,m) at a $1.272$\,GHz carrier
with $50$\,MHz bandwidth and $1024$ OFDM subcarriers; ground-truth positions
are obtained by lidar. After per-antenna phase calibration, the $1024$
subcarriers are averaged into $64$ frequency bins (PRBs) by complex averaging
over groups of $16$. Each snapshot is power-normalized and represented by
three real planes $(\log|H|,\ \cos\angle H,\ \sin\angle H)$. We then group
$T=8$ consecutive (time-ordered) snapshots into a window and assign it the
median of the corresponding ground-truth positions, yielding samples of shape
$3\times8\times64\times32$ (channels $\times$ time $\times$ frequency
$\times$ antenna). Each time snapshot is treated as a token spanning the full
frequency--antenna structure, so that masking and prediction operate along the
temporal axis. This produces $3200$ windows from $25{,}600$ snapshots.

The encoder is a Vision Transformer ($D=256$, $8$ layers, $16$ heads),
pre-trained with masked temporal latent prediction (batch $50$, EMA momentum
$0.96\!\to\!0.99$) and kept \emph{frozen} for all downstream experiments.
Localization uses a small MLP head with GELU activations, mapping the
concatenated visible embeddings $\mathbf{Z}_{\text{win}}$ to $(x,y)$; it is the
only component trained with labels (AdamW, lr $10^{-4}$, batch $40$, $500$
epochs). We use a reproducible $70/15/15$ split with a fixed seed ($480$ test
windows) and normalize target positions using training-split statistics. A
$50\%$ masking ratio is used throughout; at test time we average the head
predictions over $L{=}6$ deterministic, coverage-balanced masks (three seeds,
each contributing two complementary masks) for a mask-invariant estimate. The
supervised CNN baseline (three convolutional stages of $32/64/128$ channels with
GELU and pooling, followed by a $128{\to}256{\to}2$ MLP) is trained end-to-end
on the kept raw-CSI snapshots under the identical split, masking, and ensemble
averaging.

\subsection{Performance Metrics}
Let $\hat{\mathbf{p}}_i$ and $\mathbf{p}_i$ denote the predicted and true
$2$-D positions in meters for test sample $i$. We report the per-sample
Euclidean localization error
\begin{equation}
e_i = \lVert \hat{\mathbf{p}}_i - \mathbf{p}_i \rVert_2 ,
\end{equation}
and summarize it through the mean localization error (MLE)
\begin{equation}
\overline{e} = \frac{1}{N}\sum_{i=1}^{N} e_i .
\end{equation}
We track $\overline{e}$ on the validation set across training epochs to
monitor convergence, and report its final value on the held-out test set.
We further characterize the full error distribution via the empirical
cumulative distribution function (ECDF)
\begin{equation}
\hat{F}(d) = \frac{1}{N}\sum_{i=1}^{N} \mathbbm{1}\!\left[e_i \le d\right] ,
\end{equation}
where $\mathbbm{1}[\cdot]$ is the indicator function, equal to $1$ when its
argument is true and $0$ otherwise; hence $\hat{F}(d)$ is the fraction of
test samples localized within $d$ meters.

Because the absolute localization error in meters depends on the physical
extent and sampling of the deployment, we additionally report the relative
gain
\begin{equation}
G = \frac{\overline{e}_{\text{base}}}{\overline{e}_{\text{prop}}} ,
\end{equation}
which compares two methods evaluated on the \emph{same} scenario and thereby
isolates the contribution of the learned representation from
scenario-specific scale.

\subsection{Results}
We compare the proposed frozen-encoder localizer against a supervised
convolutional neural network (CNN) trained directly on raw CSI. On the
held-out test set, the proposed method attains an MLE of $0.42$~m,
whereas the CNN baseline reaches $0.90$~m, corresponding to a
relative gain of $G \approx 2.1\times$ (equivalently, a $53\%$ reduction in
mean error; Table~\ref{tab:main}).

\begin{table}[t]
\centering
\caption{Mean localization error on the DICHASUS test set.}
\label{tab:main}
\begin{tabular}{lc}
\toprule
Method & MLE (m) $\downarrow$ \\
\midrule
Supervised CNN (raw CSI) & $0.90$ \\
Proposed (frozen encoder + MLP head) & $\mathbf{0.42}$ \\
\bottomrule
\end{tabular}
\end{table}

\begin{figure}[t]
\centering
\includegraphics[width=0.85\columnwidth]{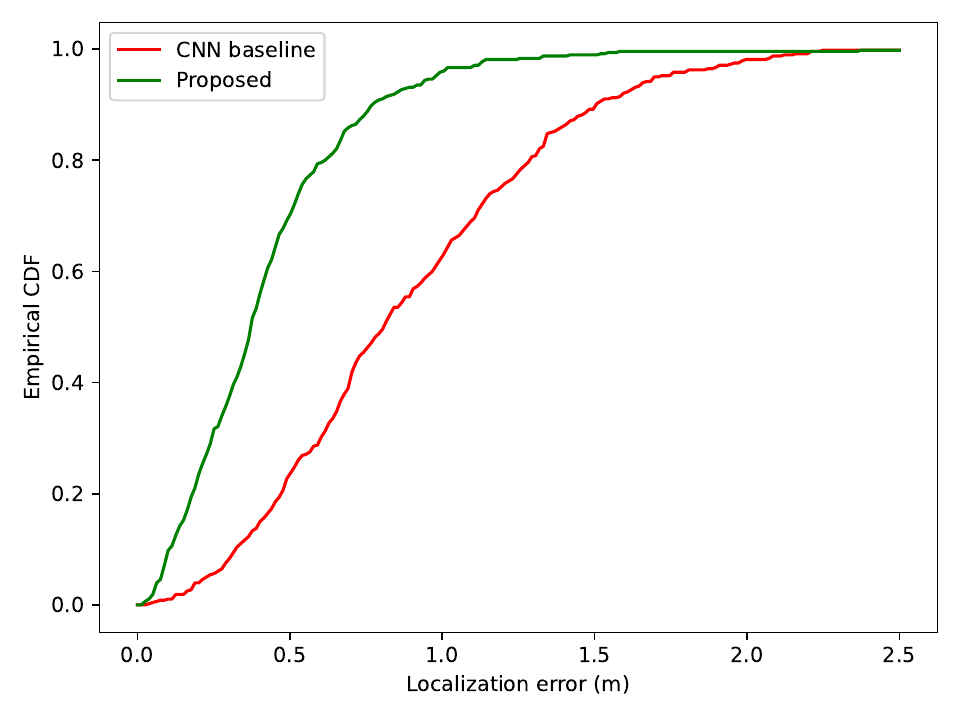}
\caption{Empirical CDF of the localization error on the test set
(CNN baseline vs.\ proposed).}
\label{fig:ecdf}
\end{figure}

\begin{figure}[t]
\centering
\includegraphics[width=0.85\columnwidth]{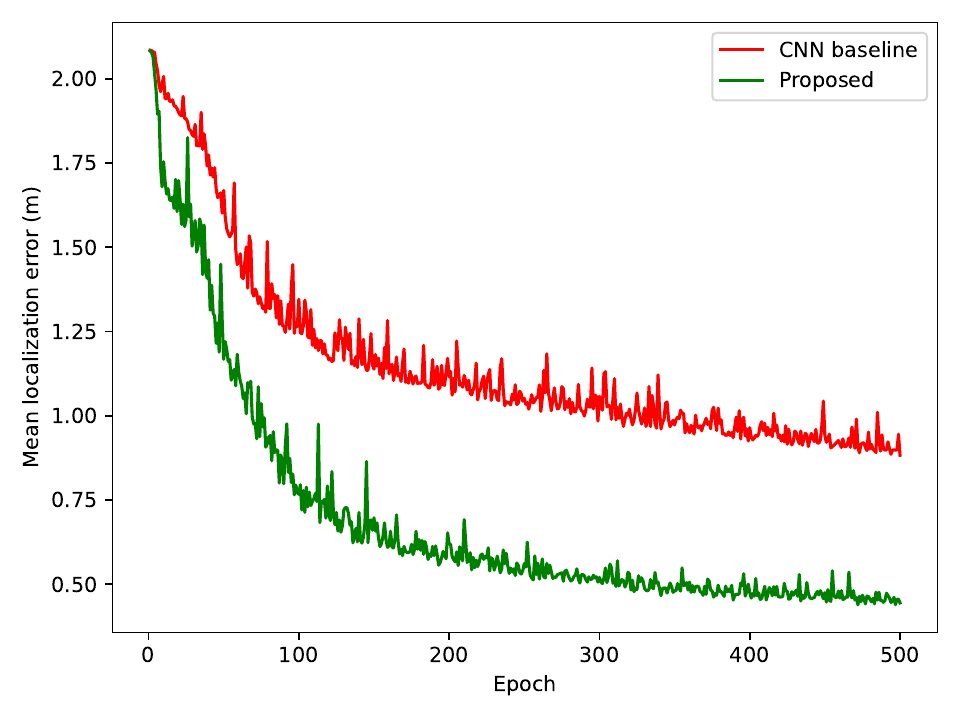}
\caption{Validation mean localization error (m) per epoch
(CNN baseline vs.\ proposed).}
\label{fig:conv}
\end{figure}

Fig.~\ref{fig:ecdf} shows the ECDF of the localization error. The curve of the
proposed system is consistently left-shifted relative to the baseline: it
localizes $70\%$ of test windows within $0.5$\,m versus only $23\%$ for the
CNN, and its $90$th-percentile error is $0.77$\,m compared to $1.51$\,m. This
indicates both a lower typical error and substantially fewer large-error
outliers.
Fig.~\ref{fig:conv} reports the validation localization error in meters over
training epochs: the error decreases rapidly and converges to a low, stable
floor. The initial error of approximately $2$~m corresponds to the
``predict-the-mean'' prior: at initialization the head outputs the
normalized origin, which de-normalizes to the mean training position
$\bar{\mathbf{p}}$, giving
$\mathrm{Err}_0 \approx \tfrac{1}{N}\sum_i \lVert \mathbf{p}_i - \bar{\mathbf{p}} \rVert_2$,
where $\bar{\mathbf{p}}$ is the mean training position. This value reflects the
spatial extent and non-uniform coverage of the DICHASUS measurement area
rather than the model itself, and is rapidly reduced as the geometry-aware
latent representation is exploited.

Finally, these results use a \emph{single-antenna} transmitter, so the spatial
signature comes solely from the $32$-element receive array; with $N_t>1$ transmit
antennas each snapshot becomes an $N_t\times N_r$ MIMO response with more spatial
degrees of freedom, which our snapshot-level tokenization absorbs with no
architectural change, so accuracy is expected to improve further.

\glsresetall
\section{Conclusion}
\label{sec:conclusion} 
In this paper, we proposed a \gls{jepa}-based self-supervised framework for
indoor localization from \gls{csi} fingerprints. Each \gls{csi} time snapshot
is treated as a token, and a Vision Transformer encoder is pre-trained by
predicting the latent representations of masked snapshots from the visible
ones, learning compact, geometry-aware representations without labeled data.
The encoder is then frozen and a \gls{lwlm} is trained on a small set of
labeled positions to estimate the user location, yielding a label-efficient
and computationally lightweight pipeline.

Experiments on the measured DICHASUS-005x dataset show that the proposed method
reduces the mean localization error from $0.90$~m to $0.42$~m relative to a
supervised \gls{cnn} trained on raw \gls{csi}---a $2.1\times$ improvement---and
yields a consistently left-shifted error distribution. Because pre-training
masks entire snapshots, the encoder operates on partial windows by
construction, which makes the pipeline naturally suited to irregular or missing
\gls{csi} measurements. These results indicate that
masked temporal latent prediction is an effective, augmentation-free route to
robust \gls{csi} representations, well aligned with \gls{isac} objectives.
Future work includes quantifying label efficiency across labeling budgets,
evaluating cross-site and cross-device generalization, and reusing the frozen
encoder for related sensing tasks such as tracking and beam selection.

\bibliographystyle{IEEEtran}
\bibliography{biblio}

\end{document}